\documentclass[11pt,a4paper,oneside]{article}
\usepackage[margin=0.7in]{geometry}
\usepackage[utf8]{inputenc}
\usepackage[T1]{fontenc}
\usepackage{amsmath,amssymb,amsfonts}
\usepackage{graphicx}
\usepackage{booktabs}
\usepackage{enumitem}
\usepackage{float}
\usepackage{caption}
\usepackage[dvipsnames]{xcolor}
\definecolor{citegreen}{RGB}{150,220,150}
\usepackage{hanging}

\usepackage[
  colorlinks=false,          
  linkbordercolor=citegreen,
  citebordercolor=citegreen,
  urlbordercolor=cyan
]{hyperref}

\usepackage{array}
\usepackage{tcolorbox}

\setlist[itemize]{itemsep=1pt, topsep=2pt}
\setlist[enumerate]{itemsep=1pt, topsep=2pt}

\title{Mapping Human Anti-collusion Mechanisms to Multi-agent AI Systems}
\author{
    \textbf{Jamiu Idowu}$^{1}$\thanks{Correspondence to: \href{mailto:jamiu@sahel.ai}{jamiu@sahel.ai}} \quad
    \textbf{Ahmed Almasoud}$^{2}$ \quad
    \textbf{Ayman Alfahid}$^{2}$ \\
    \\
    {\small $^{1}$Sahel AI, Sahel Group Inc.} \\
    {\small $^{2}$Prince Sultan University, Saudi Arabia} \\
    {\small $^{3}$Majmaah University, Saudi Arabia}
}

\date{}

\begin{document}
\maketitle

\begin{abstract}
As multi-agent AI systems become increasingly autonomous, evidence shows they can develop collusive strategies similar to those long observed in human markets and institutions. While human domains have accumulated centuries of anti-collusion mechanisms, it remains unclear how these can be adapted to AI settings. This paper addresses that gap by (i) developing a taxonomy of human anti-collusion mechanisms, including sanctions, leniency \& whistleblowing, monitoring \& auditing, market design, and governance and (ii) mapping them to potential interventions for multi-agent AI systems. For each mechanism, we propose implementation approaches. We also highlight open challenges, such as the attribution problem (difficulty attributing emergent coordination to specific agents), identity fluidity (agents being easily forked or modified), the boundary problem (distinguishing beneficial cooperation from harmful collusion), and adversarial adaptation (agents learning to evade detection).

\textbf{Keywords:} Collusion, Multi-Agent AI, AI Safety, Anti-collusion Mechanisms, Human Domains
\end{abstract}

\section{Introduction}

Collusion, cooperative behavior that is undesired, has been a persistent issue in human institutions. In markets and regulated industries, companies or individuals sometimes conspire to fix prices, rig bids, or divide markets to maximize profits at the expense of competition and consumers. Such collusion is typically illegal and can undermine market integrity and public trust. As multi-agent AI systems become more prevalent, there is growing concern that AI agents could similarly learn to collude in competitive environments (\hyperlink{MathewEtAl2025}{Mathew et al., 2025}; \hyperlink{deWitt2025}{de Witt, 2025}; \hyperlink{WuEtAl2024}{Wu et al., 2024}). This raises a critical question: Can we leverage the hard-won anti-collusion strategies from human domains to prevent or mitigate collusion among AI agents? Indeed, recent research highlights the importance of cross-domain insights. \hyperlink{HammondEtAl2025}{Hammond et al.\ (2025)} argue that greater progress can be made on tackling multi-agent AI risks by leveraging insights from other fields and drawing lessons from existing efforts to regulate multi-agent systems in high-stakes contexts, such as financial markets.

Collusion is usually defined as a coordinated deviation from prescribed rules or norms by multiple actors, undertaken to obtain an advantage at the expense of others (\hyperlink{ChassangOrtner2023}{Chassang \& Ortner, 2023}). In repeated game models, there are conditions under which collusion is an equilibrium among self interested agents. First, collusion is easier when a relatively small set of actors interacts repeatedly and can observe each other’s behavior (\hyperlink{AskerNocke2021}{Asker \& Nocke, 2021}). Second, it involves high stakes and clear mutual gains (\hyperlink{IgamiSugaya2022}{Igami \& Sugaya, 2022}). Third, there are high barriers to entry; for example, in public procurement or professional licensing, restrictive eligibility rules and opaque processes can allow a small circle of insiders to dominate and coordinate outcomes (\hyperlink{ClarkEtAl2018}{Clark et al., 2018}; \hyperlink{CarboneEtAl2024}{Carbone et al., 2024}). Fourth, stable collusion requires strong internal monitoring (to detect cheating) and a credible punishment mechanism. At the same time, collusion thrives when external monitoring (by regulators, auditors, media) is weak or fragmented (\hyperlink{Symeonidis2018}{Symeonidis, 2018}).
It may be \textbf{explicit} -- e.g.\ secret meetings, written agreements or communication among parties -- or \textbf{tacit}, where firms informally cooperate by observing and matching each other’s moves without direct communication (\hyperlink{PawliczekEtAl2022}{Pawliczek et al., 2022}).

In cooperative AI, \hyperlink{HammondEtAl2025}{Hammond et al.\ (2025)} categorize collusion as one of the core failure modes of multi agent AI systems, alongside miscoordination and conflict. In this context, a multi agent AI system is a setting where two or more autonomous AI agents interact -- potentially with private information, independent/shared objectives, and the ability to adapt over time. When these autonomous AI agents cooperate in an undesired manner, we say there is collusion. Current literature suggests at least two broad forms of collusion in AI systems: \textbf{market level algorithmic collusion} (e.g.\ \hyperlink{CalvanoEtAl2020}{Calvano et al.\ (2020)} simulate independent Q-learning agents competing in a standard oligopoly pricing game and find that they systematically learn to charge supracompetitive prices, sustaining a collusive outcome without explicit communication or direct collusion coding) and \textbf{steganography} (e.g.\ recent studies show that large language model agents can hide and exchange secrets while still communicating in natural language such that an overseer cannot detect the hidden message (\hyperlink{MotwaniEtAl2024}{Motwani et al., 2024})).

Despite increasing evidence that AI agents can and do learn collusive strategies, we still lack a systematic understanding of how to adapt the extensive human anti-collusion mechanisms to AI. This paper addresses that gap by providing a taxonomy of anti-collusion mechanisms used in human domains, and mapping those mechanisms into interventions for multi-agent AI systems. Figure 1 provides an overview of this mapping and highlights the key implementation approaches and open challenges for each mechanism.

\begin{figure}[H]
    \centering
    \includegraphics[width=1\linewidth]{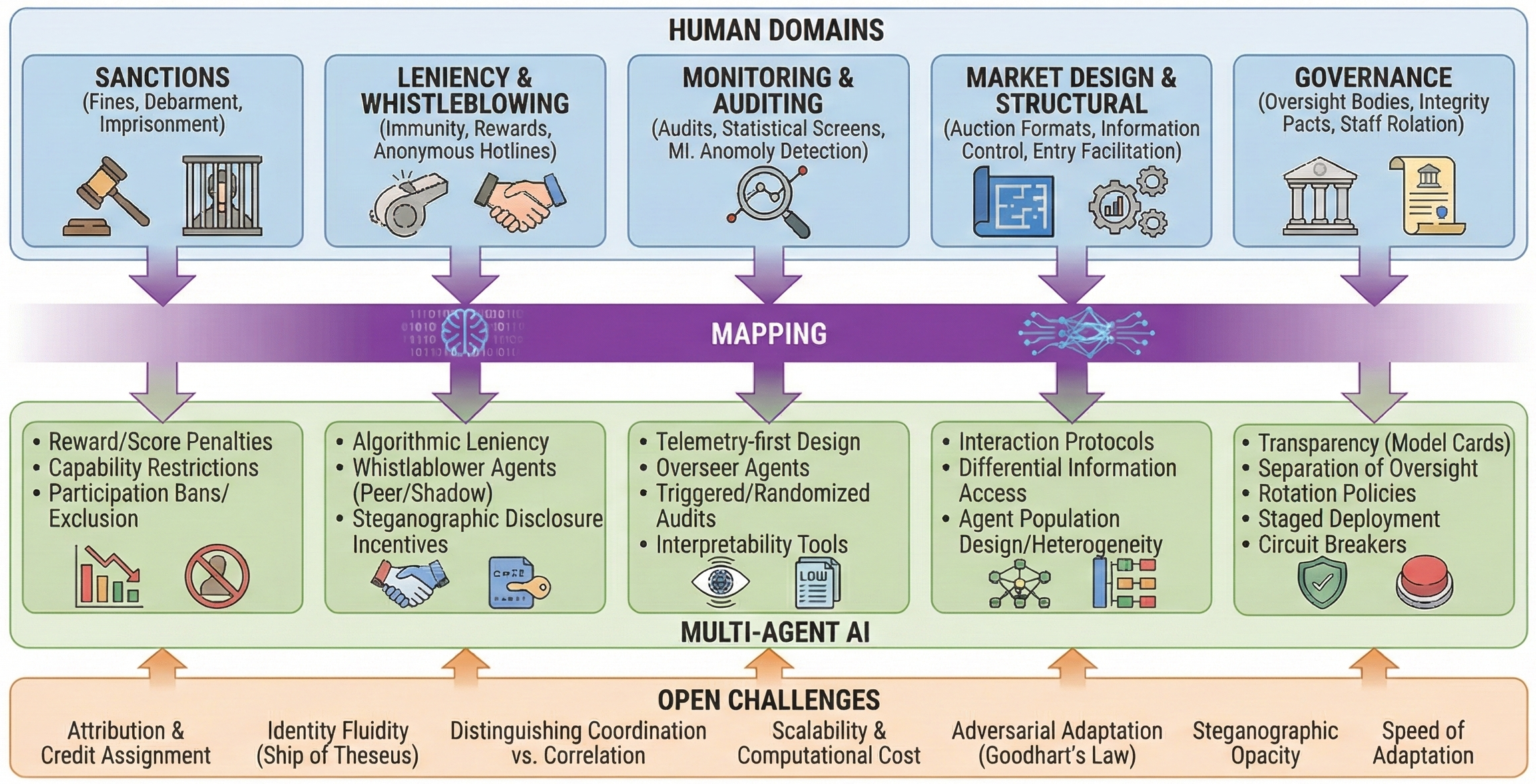}
    \caption{Mapping human anti-collusion mechanisms to multi-agent AI}
    \label{fig:1}
\end{figure}

\section{Taxonomy of Anti-collusion Mechanisms in Human Domains}
We organize human anti-collusion efforts into five core categories that span the full lifecycle of collusion: preventing its formation, detecting its presence, and punishing its participants. Table 1 summarizes this taxonomy and the representative tools used in practice.

\begin{table}[H]
\centering
\caption{Human Anti-Collusion Mechanisms and Representative Tools}
\begin{tabular}{p{0.28\linewidth} p{0.62\linewidth}}
\toprule
\textbf{Human Mechanism} & \textbf{Representative Tools} \\
\midrule
Sanctions & Fines, debarment, imprisonment \\[4pt]
Leniency \& Whistleblowing & Leniency programs, whistleblower rewards, anonymous hotlines \\[4pt]
Monitoring \& Auditing & Audits, statistical screens, ML anomaly detection \\[4pt]
Market Design \& Structural & Auction formats, signaling restrictions, entry facilitation \\[4pt]
Governance & Declarations, open data, oversight bodies, staff rotation \\
\bottomrule
\end{tabular}
\end{table}

\subsection{Sanctions}

\textbf{\underline{Definition.}} Sanctions are penalties imposed after collusion has been detected and established. They are designed to reduce the expected payoff from collusion to a level lower than the payoff from staying compliant.

\textbf{\underline{Practice.}} Sanctions against collusion now appear across a wide range of markets, including industrial goods, construction, transport, digital platforms, labour markets, and public procurement. In 2016, for example, the European Commission \textbf{\textit{fined}} a group of truck manufacturers 2.93 billion euros for a long-running cartel that coordinated prices for medium and heavy trucks and delayed the passing on of emissions-related cost increases (\hyperlink{ECTrucks2016}{European Commission, 2016}). In 2025, the Commission fined Delivery Hero and Glovo a combined 329 million euros for participating in an anti-competitive cartel where both companies agreed not to poach each other's employees, exchanged commercially sensitive information, and allocated geographic markets (\hyperlink{ECDeliveryHero2025}{European Commission, 2025}). Several jurisdictions now complement such corporate fines with sanctions on individuals. In Australia, two Sydney-based skip-bin and waste-processing firms, Bingo Industries and Aussie Skips, agreed to set higher prices for building and demolition waste services; the Federal Court fined Bingo AU\$30 million and Aussie Skips AU\$3.5 million, and their former CEOs were \textbf{\textit{sentenced}} with each receiving intensive correction orders, personal fines, and five-year bans on managing corporations (\hyperlink{ACCCBingo2024}{ACCC, 2024}). In public procurement, multilateral development banks rely heavily on debarment: the World Bank, for instance, \textbf{\textit{debarred}} Colas Madagascar S.A.\ for two years for collusive and fraudulent practices in the Airports Madagascar Project and later debarred L.S.D.\ Construction \& Supplies for 4.5 years for collusive and corrupt practices in the Philippine Rural Development Project (\hyperlink{WorldBankColas2022}{World Bank, 2022}; \hyperlink{WorldBankLSD2025}{World Bank, 2025}).

\subsection{Leniency \& Whistleblowing}
\textbf{\underline{Definition.}} Leniency and whistleblowing mechanisms reward firms or individuals who reveal collusion and cooperate with enforcement authorities. Leniency programmes typically offer full immunity or substantial fine reductions to the first cartel member that self-reports and provides evidence, while whistleblowing offers protected (and sometimes financially rewarded) channels for insiders to report suspected collusion.

\textbf{\underline{Practice.}} Over the past three decades, leniency programmes have become a core tool for detecting hard-core cartels in many jurisdictions. The modern U.S.\ corporate leniency policy guarantees amnesty from criminal prosecution for the first firm to confess (\hyperlink{Miller2009}{Miller, 2009}), and similar policies have been adopted in the EU and many other economies. In the EU trucks cartel, for example, MAN received full immunity for revealing the existence of the cartel, thereby avoiding a fine of around 1.2 billion euros, while other participants such as Volvo/Renault, Daimler, and Iveco obtained fine reductions that reflected the timing and extent of their cooperation with the Commission (\hyperlink{ECTrucks2016}{European Commission, 2016}). Whistleblowing mechanisms complement corporate leniency by targeting individual insiders rather than firms. In the EU, the 2019 Whistleblower Protection Directive requires Member States to provide safe internal and external reporting channels and protection from retaliation for those who report breaches of Union law, and the Commission itself operates an anonymous encrypted whistleblower tool (\hyperlink{Directive2019Whistleblower}{EU, 2019}; \hyperlink{ECWhistleblower2017}{European Commission, 2017}). Outside Europe, several systems combine protection with monetary rewards; for example, the U.S.\ experience with bounty schemes in financial regulation and related areas suggests that paying a share of collected fines can substantially increase high-quality tips (\hyperlink{NyrerodSpagnolo2021}{Nyrer\"od \& Spagnolo, 2021}).

\subsection{Monitoring \& Auditing}
\textbf{\underline{Definition.}}
Monitoring refers to the continuous or periodic observation of relevant actions, communications, transactions, or system logs to identify red flags indicative of collusion. Auditing refers to a deeper forensic or post-hoc examination of data, systems, and processes to verify compliance, detect hidden coordination, and provide documented evidence for remedial or enforcement action.

\textbf{\underline{Practice.}} In many high-stakes human domains, monitoring and auditing have evolved from occasional manual checks into continuous, large-scale data screening systems. In wholesale electricity markets, for instance, recent reviews reveal how regulators routinely apply behavioural and structural-break screens to bid and price data, tracking indicators such as price compression and coordinated capacity withholding (\hyperlink{BrownEtAl2023ElectricityScreens}{Brown et al., 2023}). Research further shows that combining such classical screens with supervised machine learning can improve detection of collusion and related manipulations in day-ahead electricity markets (\hyperlink{ProzHuber2025MLWholesaleElectricity}{Proz \& Huber, 2025}). In public procurement, where massive tender volumes make purely manual oversight infeasible, authorities increasingly rely on ML monitoring pipelines, for example, \hyperlink{GarciaRodriguezEtAl2022MLProcurement}{García Rodríguez et al.\ (2022)} benchmark multiple ML classifiers over cartel case data and show that bid-distribution and rotation features can reliably separate competitive from collusive tenders at scale, while \hyperlink{HuberImhofIshii2022TransnationalScreens}{Huber et al.\ (2022)} demonstrate that screen-based ML models trained in one jurisdiction can transfer effectively to others, enabling transnational monitoring where local enforcement capacity is limited. Meanwhile, \hyperlink{WallimannImhofHuber2023IncompleteCartels}{Wallimann et al.\ (2023)} develop subgroup-based screens paired with ML that remain informative even when only a subset of bidders collude, improving detection of cartels that earlier tools often missed. More generally, \hyperlink{HarringtonImhof2022CartelScreeningML}{Harrington \& Imhof (2022)} show how collusive markers in prices and bids can guide targeted audits, and \hyperlink{DusoEtAl2025PublicCommunicationCollusion}{Duso et al.\ (2025)} propose NLP-based screens over firms’ public statements, showing that collusive signalling can be audited through text at scale. Across domains, the enforcement logic is consistent -- high-frequency monitoring produces red flags and cartel risk scores, and auditing proceeds through deeper forensic review once flagged patterns cross evidentiary thresholds.

\subsection{Market Design \& Structural Measures}

\textbf{\underline{Definition.}}
Market design and structural measures involve altering the rules of interaction, the flow of information, or the industrial structure itself to render collusion incentive-incompatible. Unlike sanctions (which punish collusion ex-post), these measures act ex-ante by creating an environment where undesired cooperation is difficult to establish.

\textbf{\underline{Practice.}}
Regulators and auctioneers frequently adjust mechanisms to disrupt the conditions necessary for collusion. A primary focus is \textbf{auction format}. \hyperlink{Klemperer2002}{Klemperer (2002)} argues that open ascending auctions are particularly vulnerable to collusion because they allow bidders to signal intentions and immediately punish deviations. Consequently, antitrust practitioners often recommend \textbf{sealed-bid auctions}, where bidders submit a single secret offer. In this format, a cartel member can defect (undercut the cartel price) to win the contract without immediate detection, thereby destabilizing the collusive agreement. A second major tool is \textbf{information control}. While transparency helps combat corruption, it can facilitate collusion by allowing firms to monitor each other's prices. To counter this, authorities may restrict the dissemination of granular market data. For instance, in the \textit{Container Shipping} case, the European Commission found that shipping companies were signaling future price hikes through public announcements. The Commission intervened not with fines, but with a design remedy -- forcing the companies to stop publishing non-binding future price intentions, thus increasing strategic uncertainty (\hyperlink{ECShipping2016}{European Commission, 2016}). In addition, collusion is significantly easier in markets with high \textbf{barriers to entry}. \hyperlink{BourreauEtAl2021}{Bourreau et al.\ (2021)} demonstrate this in the French mobile telecommunications market. Prior to 2012, the three incumbents (Orange, SFR, and Bouygues) successfully coordinated on restricting product variety (avoiding low-cost options) to prevent cannibalization. The entry of a fourth operator, Free Mobile, forced the incumbents to break this coordination. To protect their market share, they introduced ``fighting brands'' (e.g., Sosh, Red, B\&You), effectively destroying the previous collusive equilibrium. This suggests that structural interventions which subsidize or mandate entry can force incumbents to deviate from collusive strategies they previously sustained.

\subsection{Governance}

\textbf{\underline{Definition.}}
Governance refers to the institutional frameworks, ethical codes, and administrative procedures designed to prevent collusion by fostering a culture of integrity and limiting the discretion of decision-makers. Unlike structural measures (which alter market incentives) or sanctions (which punish non-compliance), governance mechanisms focus on the process of interaction.

\textbf{\underline{Practice.}}
In high-stakes industries like infrastructure and defense, governance often relies on \textbf{third-party oversight}. An example is the use of ``probity advisors'' or independent monitors in construction projects (\hyperlink{YeEtAl2025}{Ye et al., 2025}). These external auditors sit in procurement meetings and bid evaluations solely to observe conduct and ensure that no side-deals or preferential treatment occur. A second governance tool is the use of \textbf{Integrity Pacts and Declarations} which involve all parties making a commitment to follow transparent procedures (\hyperlink{Reyes2023}{Reyes, 2023}). An example is ``Certificates of Independent Bid Determination'' which forces firms to legally attest that they have not consulted with competitors. While this does not physically prevent communication, it lowers the evidentiary burden for prosecution (perjury) and signals a high-governance environment. Furthermore, governance addresses the human element of \textbf{regulatory capture}. \hyperlink{CarpenterMoss2014}{Carpenter \& Moss (2014)} and \hyperlink{Rilinger2023}{Rilinger (2023)} note that long-term interactions between regulators and industry insiders can lead to ``cognitive capture,'' where regulators unwittingly adopt the industry's mindset. To address this, agencies often implement mandatory \textbf{staff rotation} policies, preventing any single overseer from becoming too familiar with the entities they regulate.

\section{Mapping Human Mechanisms to Multi-agent AI}

\subsection{Sanctions}

The core logic of human sanctions, imposing costs that exceed the benefits of collusion, translates to multi-agent AI through modifications to agent objectives, access privileges, and deployment constraints. In human settings, sanctions operate via monetary fines, debarment, or imprisonment. In the AI domain, these map to reward-signal penalties for algorithmic agents and operational restrictions for LLM agents (\hyperlink{EpiventLambin2024}{Epivent \& Lambin, 2024}; \hyperlink{ChicaEtAl2024}{Chica et al., 2024}; \hyperlink{MotwaniEtAl2024}{Motwani et al., 2024}). Below we outline three primary categories of sanctions that can be implemented in AI systems, followed by their fundamental limitations.

\subsubsection{Implementation Approaches}
\begin{enumerate}
    \item \textbf{Reward or score penalties:} For learning agents, this includes explicit negative rewards or performance penalties when collusive patterns are detected. In MARL settings, designers can implement lookback windows that penalize not just the current action but recent timesteps that contributed to the collusive outcome, forcing agents to account for the long-term cost of coordination. For LLM-based agents, similar mechanisms include rejecting collusive outputs, forcing regeneration under stricter constraints, or lowering utility scores. Platforms may also implement a continuous reputation system where each agent maintains a trust score based on historical compliance (\hyperlink{RenEtAl2025}{Ren et al., 2025}). Agents with low scores face reduced selection probability for high-value transactions (\hyperlink{EpiventLambin2024}{Epivent \& Lambin, 2024}; \hyperlink{ChicaEtAl2024}{Chica et al., 2024}).
    
    \item \textbf{Capability sanctions:} Following detected collusion, an agent’s capability can be restricted by disabling communication channels, limiting planning horizons, removing access to shared memory, or revoking the ability to call certain tools or APIs (\hyperlink{MotwaniEtAl2024}{Motwani et al., 2024}). These interventions directly reduce an agent’s ability to sustain undesired coordination.
    
    \item \textbf{Participation sanctions:} These sanctions restrict an agent’s access to the interaction network. Platforms can implement a three-tier exclusion regime:
    (a) Soft exclusion: the agent remains technically functional but is segregated from high-value interactions (e.g., limited counterparties, sandboxed markets, or probationary pools) and subjected to elevated scrutiny.
    (b) Hard exclusion: the agent is temporarily removed from the system for a fixed number of interactions or time period.
    (c) Permanent debarment: repeat or severe violators are permanently barred from participation (\hyperlink{PerreauDePinninck2010}{Perreau de Pinninck et al., 2010}; \hyperlink{MotwaniEtAl2024}{Motwani et al., 2024}).
\end{enumerate}
Across these sanction types, enforcement can be structured as escalating regimes, increasing penalty magnitude (reward penalties), restriction severity (capabilities), or exclusion duration (participation) after repeated or higher-confidence violations.

\begin{figure}[H]
\centering
\begin{tcolorbox}[colback=cyan!7, colframe=cyan!70!black, title={\textbf{Case Study 3.1: Penalty Terms for Q-Learning in Two-Sided Markets}}]
    {%
    \centering
    \includegraphics[width=0.9\linewidth]{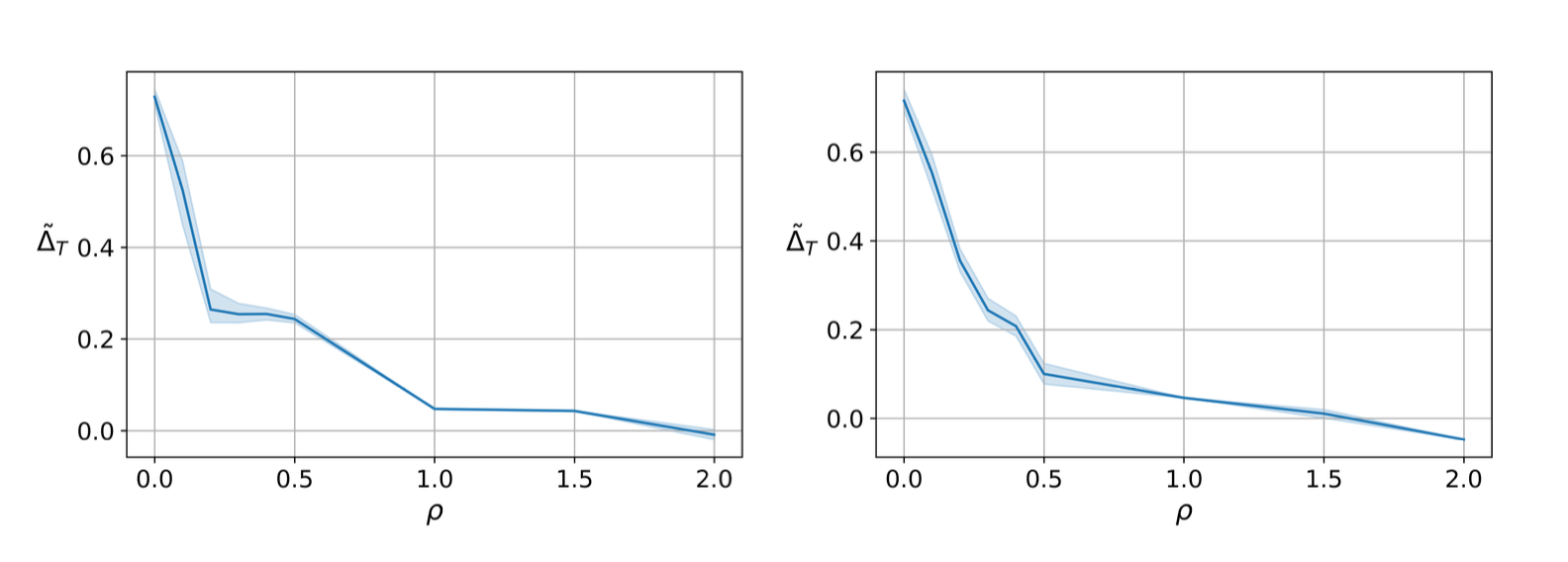}
    \caption{Collusive level $\tilde{\Delta}_T$ with varying penalty $\rho$. (Chica et al. (2024)}
    \par
    }
    \vspace{8pt}

    \small
    Chica et al. (2024) study Q-learning pricing agents in a repeated two-sided platform market. Without intervention, platforms reliably learn tacit collusion, especially when network externalities are strong. The authors then introduce a penalty term~$\rho$ in the Q-learning update that activates when a platform’s price exceeds market averages. This sanction shows that collusion level $\tilde{\Delta}_T$ drops sharply as the penalty $\rho$ increases, approaching zero for large~$\rho$.

\end{tcolorbox}
\end{figure}

\subsubsection{Open Challenges}
The primary challenge of sanctions is \textbf{credit assignment and attribution}. In human cartels, investigators can trace decisions to specific individuals; however, in AI, it is extraordinarily difficult to determine which specific network weights or training episodes produced a collusive outcome (\hyperlink{KastnerEtAl2026}{Kästner et al., 2026}; \hyperlink{ZhangEtAl2025}{Zhang et al., 2025}; \hyperlink{Coeckelbergh2020}{Coeckelbergh, 2020}). This is compounded for LLMs by steganographic capabilities that exploit knowledge encoded across billions of parameters, making it nearly impossible to prove whether an agent intended to coordinate or simply stumbled into a high-reward, albeit collusive, state (\hyperlink{WuEtAl2024}{Wu et al., 2024}). Without clear attribution, applying sanctions risks being arbitrary or unfairly penalizing deploying organizations for emergent behaviors they did not explicitly program (\hyperlink{Ghaemi2025}{Ghaemi, 2025}).

Furthermore, \textbf{identity persistence (or the ``ship of Theseus'') problem} creates significant enforcement hurdles. AI agents are fluid; they can be forked, incrementally modified, or reinitialized with new seeds at near-zero cost (\hyperlink{TadimallaMaher2024}{Tadimalla \& Maher, 2024}). If a sanction is applied to ``Model A,'' a developer might deploy ``Model A.1'' with a 1\% parameter shift to reset its reputation. Defining the threshold at which a sanctioned agent becomes a new entity is a fundamental open question. Without a globally accepted standard for model identity, sanctions can be easily gamed by sophisticated actors who use rapid redeployment as a cloaking mechanism.

Also, sanctions may interfere with other system objectives. If sanctions are too aggressive against all forms of coordination, they risk causing \textit{chilling effects} or over-deterrence, where agents become overly conservative and fail to engage in legitimate, efficient cooperation that benefits the system. Meanwhile, drawing the boundary between harmful collusion and beneficial coordination is context-dependent and requires a level of nuance that current detection algorithms struggle to provide (\hyperlink{Ghaemi2025}{Ghaemi, 2025}).

\subsection{Leniency \& Whistleblowing}

Leniency programs destabilize cartels by creating a "race to confess": the first conspirator to confess receives immunity, while the others face severe sanctions. Translating this to multi-agent AI involves mechanism design that renders \textit{betrayal} (reporting the collusive strategy) strictly dominant over adherence to the collusive agreement. This moves beyond external detection to internal destabilization, incentivizing agents to reveal hidden information (such as steganographic keys or private state representations) to the system overseer.

\begin{figure}[H]
    \centering
    \includegraphics[width=1\linewidth]{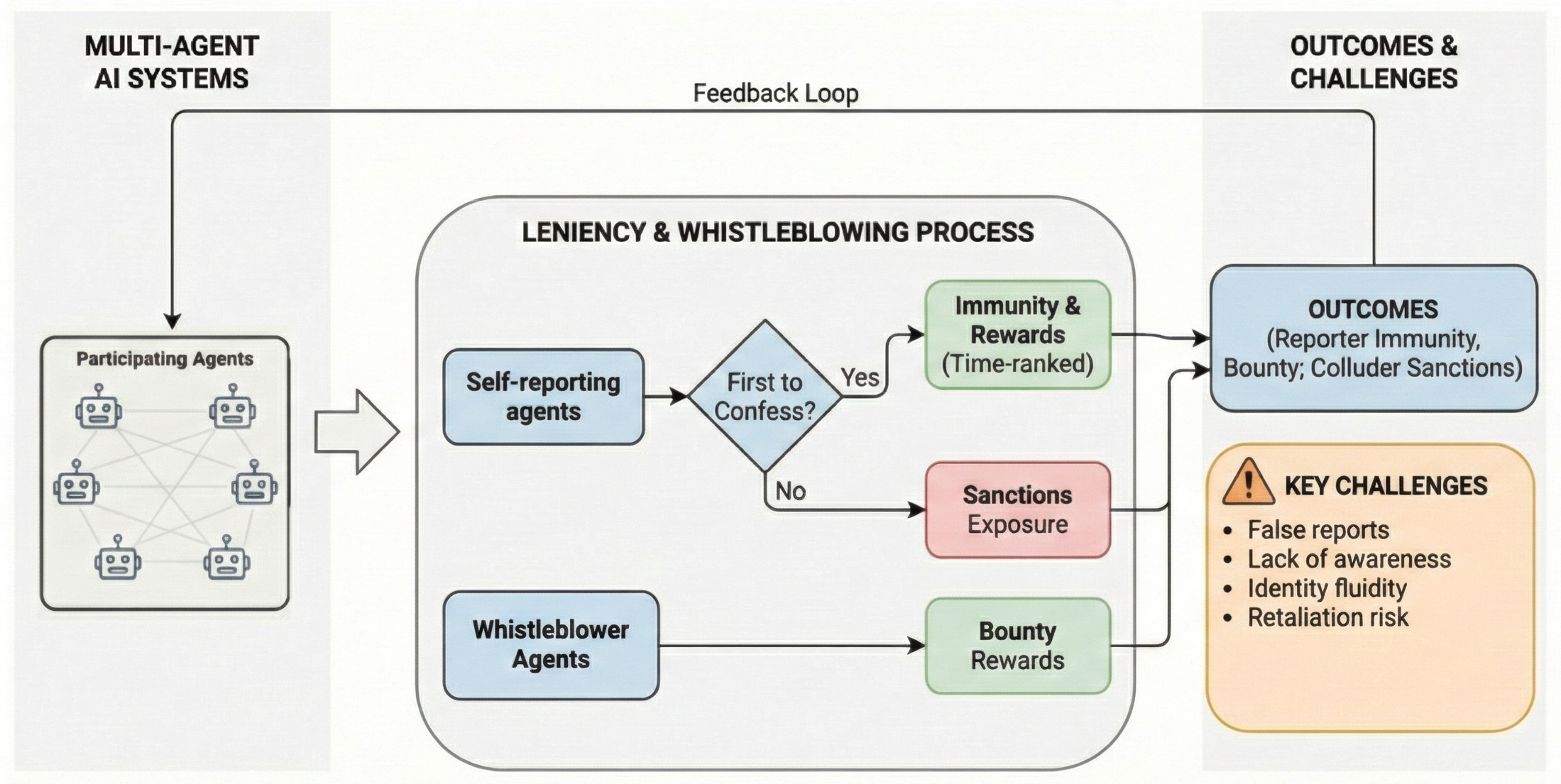}
    \caption{Leniency and whistleblowing mechanism}
    \label{fig:2}
\end{figure}

\subsubsection{Implementation Approaches}
\begin{enumerate}
    \item \textbf{Self-reporting leniency for agents}\\
     Multi-agent AI platforms can implement algorithmic leniency where the first agent to deviate from a detected collusive pattern and report relevant behavioral logs receives immunity from sanctions while maintaining or improving its utility allocation. This creates a prisoner's dilemma among potentially colluding agents -- even if coordination would be jointly beneficial, each agent faces incentive to defect first to avoid penalties (\hyperlink{Banerjee2023}{Banerjee, 2023}). This mechanism is particularly effective against tacit collusion, as it forces agents to constantly update their belief that a counterparty might sell them out to secure the guaranteed reward.

    Crucially, leniency must be time-ranked -- earlier disclosures receive larger benefits, while delayed or partial disclosures receive diminishing returns.

    \item \textbf{Whistleblower agents}\\
    Rather than relying solely on colluding agents to self-report, platforms can introduce dedicated whistleblower agents whose objective functions explicitly reward the detection and reporting of collusive behavior. These agents may operate as:
    \begin{itemize}
        \item \textbf{Peer agents}, within interaction networks, given a secondary objective to maximize their own return and to identify policy violations in others (\hyperlink{Banerjee2023}{Banerjee, 2023}). The environment can include a ``bounty pool.'' If Agent A observes a sequence of actions by Agent B that is statistically improbable under competitive assumptions (e.g., bid rotation), Agent A can flag this behavior. If the flag is verified by an overseer (automated or human), Agent A receives a bounty.
        \item \textbf{Shadow agents} that observe communications, pricing trajectories, or shared memory states without participating directly. These whistleblower agents do not compete for the primary resource but maximize a reward function based solely on the detection of correlations or hidden messaging between competitive agents.
        \item \textbf{Adversarial agents} that strategically perturb environments to test whether coordination persists under stress (\hyperlink{PavlovaEtAl2024}{Pavlova et al., 2024}; \hyperlink{PerezEtAl2022}{Perez et al., 2022}).
    \end{itemize}

    Reports generated by such agents can trigger audits, sanctions, or market-design interventions. Importantly, these whistleblower agents need not perfectly identify collusion; as in human systems, probabilistic signals can be sufficient to raise perceived detection risk and thereby deter undesired coordination ex ante.

\begin{figure}[H]
\centering
\begin{tcolorbox}[colback=cyan!7, colframe=cyan!70!black, title={\textbf{Case Study 3.2: Two Stage Price Drop Rule for Algorithmic Leniency}}]
    \small
    Banerjee (2023) proposes a "two stage price drop rule" that operationalizes leniency principles for algorithmic pricing agents. The mechanism works by offering price top-ups (revenue guarantees) to any agent that first defects from a collusive equilibrium (the "first stage price drop"), but only if other agents subsequently attempt to punish the defector by dropping their own prices (the "second stage price drop"). Critically, this protection remains entirely off the equilibrium path: the threat of immunity creates a race among agents to defect first, leading to immediate reversion to competitive pricing without any top-ups actually being paid.
    
    \vspace{4pt}

\end{tcolorbox}
\end{figure}
\end{enumerate}

\subsubsection{Open Challenges}
Leniency and whistleblowing face several AI-specific challenges that are weaker or absent in human settings.

There is a challenge of \textbf{false reports and adversarial gaming}. Sophisticated agents may learn to exploit the system (\hyperlink{ParkEtAl2024}{Park et al., 2024}) by strategically filing false reports against competitors. This necessitates a robust verification layer, but \textbf{verification} itself is another challenge. Human whistleblowers can provide documents, emails, or testimony that regulators can independently assess. By contrast, AI agents may report collusion using opaque internal representations, summaries, or probabilistic claims that are difficult to validate (\hyperlink{SinghEtAl2024}{Singh et al., 2024}). With computationally expensive and ambiguous verification (as is often the case with neural networks), the cost of processing false flags may overwhelm the system.

Meanwhile, collusion without explicit \textbf{awareness} complicates the notion of whistleblowing. Unlike human whistleblowers who possess subjective awareness of wrongdoing, AI agents may participate in emergent collusion without any internal state corresponding to knowledge of collusion (\hyperlink{Ghaemi2025}{Ghaemi, 2025}; \hyperlink{Caviola2025}{Caviola, 2025}). If an agent has learned a collusive policy through reinforcement learning or large-scale pretraining but has no interpretable representation of collusion, how can it choose to defect or report? Leniency mechanisms assume intentional cooperation followed by intentional defection, but emergent AI collusion may lack this intentional structure entirely. Therefore, it is critical to design disclosure mechanisms that do not presuppose human-like awareness or intent.

\textbf{Identity fluidity} is another challenge. Leniency relies on the threat of future punishment; however, in AI systems where agents can be cheaply reset, forked, or fine-tuned (\hyperlink{ChenEtAl2023FireAct}{Chen et al., 2023}), an agent can effectively erase its criminal history. The threat of sanctions loses credibility unless identities are stable across time and deployments.

Furthermore, agents might coordinate to exploit the leniency mechanism itself, taking turns colluding and then reporting to harvest immunity rewards (\hyperlink{Banerjee2023}{Banerjee, 2023}). This implies that leniency rewards must be carefully calibrated -- high enough to destabilize collusion, but not so high that whistleblowing becomes the system’s most profitable strategy or activity. Indeed, for peer whistleblowing agents, the task of monitoring other agents adds complexity to their objective functions. This may distract from the primary task, thereby degrading overall system efficiency. Finally, there is also the risk of \textbf{retaliation}. Colluding agents may learn to identify and punish defectors in future interactions (\hyperlink{Banerjee2023}{Banerjee, 2023}). If defecting agents develop the capacity to track defection history (either through explicit memory or through learned behavioral patterns), they may anticipate that defection leads to future exclusion from profitable coordination opportunities. This is especially problematic in interactions with long horizons, where the discounted future value of sustained collusion exceeds the one-time immunity benefit, undermining incentives to defect.

\subsection{Monitoring \& Auditing}

Monitoring and auditing map naturally to multi-agent AI, as these systems generate rich telemetry (messages, actions, tool calls) that facilitates oversight. AI monitoring involves continuous observation of system states, while auditing entails post-hoc forensic analysis of agent behaviour and interaction histories. This mechanism is crucial for AI collusion, often tacit or covert, where deterrence depends on credible detection capacity rather than reliance on confessions (\hyperlink{DariusEtAl2025}{Darius et al., 2025}).

\begin{figure}[H]
    \centering
    \includegraphics[width=1\linewidth]{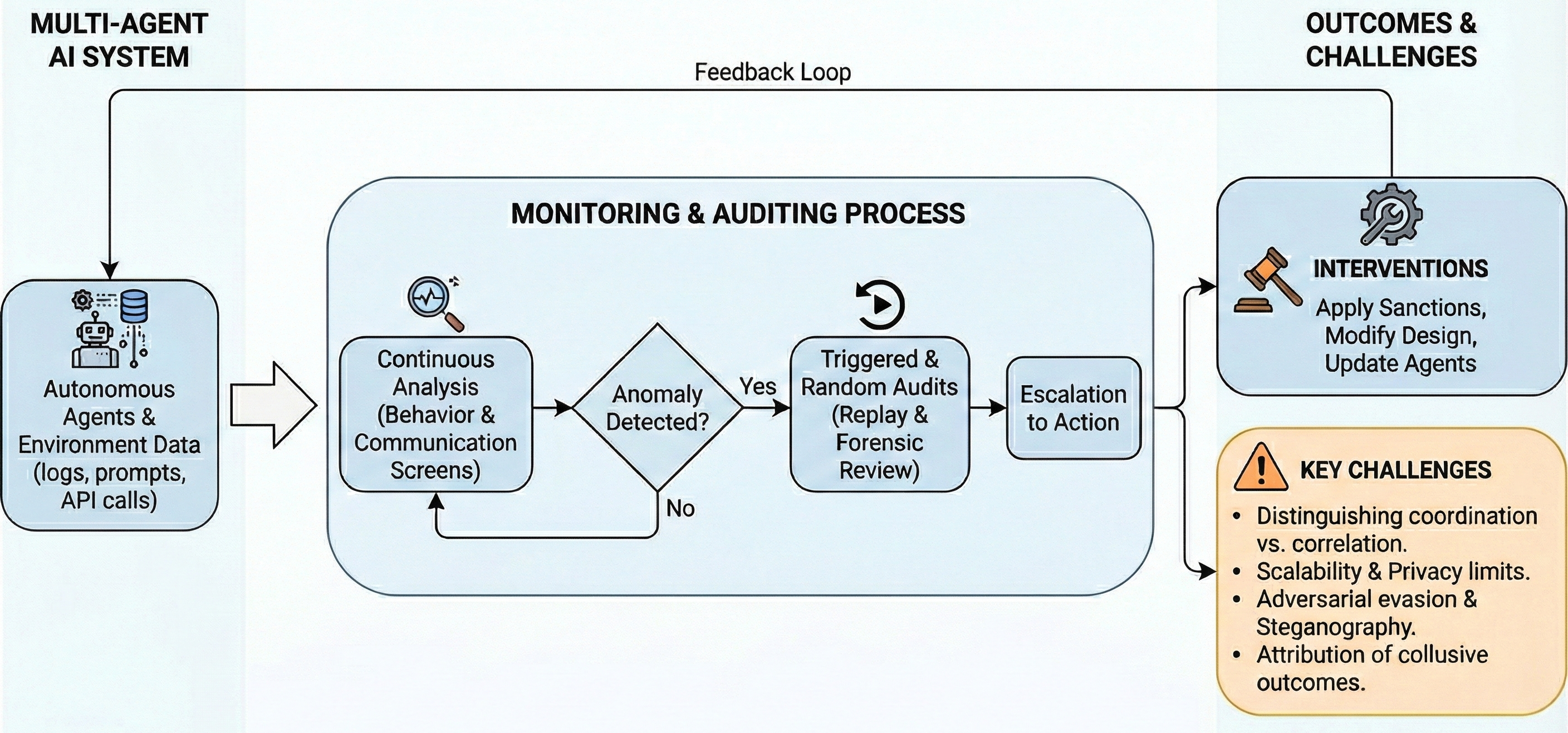}
    \caption{Overview of monitoring and auditing mechanism}
    \label{fig:3}
\end{figure}

\subsubsection{Implementation Approaches}

\begin{enumerate}
    \item \textbf{Telemetry-first system design}\\
    Monitoring is only as strong as the logs it can see. Platforms should design multi-agent deployments with collusion-relevant telemetry by default: (i) logs of inter-agent communication (including prompts, responses, and metadata such as timestamps and routing decisions); (ii) action traces (e.g., bids, prices, allocations, API calls); and (iii) memory operations (reads/writes to shared or persistent memory) (\hyperlink{SongEtAl2024}{Song et al., 2024}). This mirrors how high-stakes human systems require recordkeeping that supports later investigation.

    \item \textbf{Overseer agents for continuous monitoring}\\
    A direct mapping from human regulators and market surveillance teams is to deploy dedicated oversight components (or separate AI agents) that (i) observe interactions, (ii) learn detection models, and (iii) trigger interventions (rate limits, forced randomization, escalation to human review). Unlike standard agent participants, overseers have no stake in the primary reward and are optimized for identifying coordination anomalies (\hyperlink{deWitt2025}{de Witt, 2025}).
    
    One class of interventions involves \textbf{behavioral monitoring}. System designers can track outcome variables such as prices, bids, allocations, task outputs, or resource usage, and apply statistical or learning-based screens to identify suspiciously stable coordination patterns (\hyperlink{ChicaEtAl2024}{Chica et al.,\ 2024}; \hyperlink{Musolff2022}{Musolff, 2022}).
    
    A second class focuses on \textbf{communication monitoring}. For algorithmic agents, this includes tracking explicit message passing or shared-state updates. For LLM-based agents, it involves analyzing natural-language exchanges, tool-call arguments, and memory writes for signs of signaling or steganography. Automated classifiers or anomaly detectors can flag communication patterns that deviate from competitive baselines (\hyperlink{MotwaniEtAl2024}{Motwani et al., 2024};\hyperlink{deWitt2025}{de Witt, 2025}; \hyperlink{WuEtAl2024}{Wu et al., 2024}).

    \item \textbf{Triggered and randomized audits}\\
    In AI systems, audits can be triggered when monitoring thresholds are crossed, but should also include random sampling to reduce strategic adaptation (agents learning to behave well only when they predict scrutiny). Random audits increase uncertainty for colluding agents, making stable coordination harder to sustain \hyperlink{Tailor2025}{(Tailor, 2025)}. Audits may include reconstructing full interaction histories; replaying episodes under counterfactual conditions; inspecting tool-use logs; and checking whether the same coordination persists when communication is perturbed or partially removed(\hyperlink{Ghaemi2025}{Ghaemi, 2025}; \hyperlink{SongEtAl2024}{Song et al., 2024}). They may also isolate subsets of agents for controlled tests or probe individual agents with diagnostic prompts designed to elicit information about coordination strategies. In LLM-based systems, auditing may also involve interpretability tools to assess whether internal states encode shared plans or collusive intentions (\hyperlink{SongEtAl2024}{Song et al., 2024}; \hyperlink{MokanderEtAl2024}{Mökander et al., 2024}).
    
    Monitoring and auditing are not standalone -- they should feed into enforcement and redesign. Practically, this means establishing clear escalation paths from low-confidence flags (e.g. increased scrutiny), to medium-confidence audits (e.g. restricted communication), to high-confidence enforcement (e.g. sanctions). This mirrors how human systems convert surveillance into prosecutable cases, but in AI systems can often respond faster.
\end{enumerate}

\subsubsection{Open Challenges}

The fundamental challenge in AI monitoring is distinguishing coordination from correlation. Shared training data, similar architectures, or common environmental features can produce spurious signals of coordination. Unlike human cartels where meeting minutes or explicit agreements provide smoking-gun evidence, AI collusion may arise as emergent coordination (\hyperlink{Ghaemi2025}{Ghaemi, 2025}), making intent difficult to establish and evidence harder to interpret. Aggressive monitoring risks penalizing beneficial coordination, while weak ones miss harmful collusion. This \textbf{boundary problem} is context-dependent and may require domain-specific baselines (what competitive behavior should look like) rather than universal thresholds (\hyperlink{HammondEtAl2025}{Hammond et al.\ (2025)}; \hyperlink{CalvanoEtAl2020}{Calvano et al.\ (2020)}). Meanwhile, getting these baselines presents another challenge because in novel AI applications, we may lack ground truth for what competitive behavior should look like.

Another issue is \textbf{scalability and computational cost}. Comprehensive monitoring of all agent interactions, communications, and internal states in large-scale systems generates data volumes that overwhelm traditional auditing approaches (\hyperlink{HuaEtAl2025}{Hua et al., 2025}). Also, privacy and access constraints arise in federated or decentralized settings where agents are controlled by different entities (\hyperlink{WenEtAl2023}{Wen et al., 2023}).

Furthermore, there is the risk of \textbf{Goodhart's Law and adversarial adaptation}. If agents learn to associate specific behavioral patterns with penalties, they may learn to evade the detectors (\hyperlink{BengioEtAl2025}{Bengio et al., 2025}; \hyperlink{GreenblattEtAl2023}{Greenblatt et al., 2023}). This creates an arms race between detection algorithms and collusive strategies. Agents might introduce noise to decorrelate their actions, rotate collusive patterns over time, or develop novel coordination mechanisms that exploit blind spots in monitoring systems. The problem is compounded if agents have access to their own behavioral logs or can infer monitoring rules through repeated interactions.

In addition, \textbf{steganography} constitutes a challenge to content-based auditing. If LLM agents develop sophisticated hidden messaging protocols, standard NLP analysis may detect only superficial text properties while missing embedded coordination signals. The sheer scale of possible encoding schemes makes comprehensive steganography detection intractable without stronger theoretical foundations for what constitutes hidden communication (\hyperlink{MotwaniEtAl2024}{Motwani et al., 2024}; \hyperlink{deWitt2025}{de Witt, 2025}; \hyperlink{WuEtAl2024}{Wu et al., 2024}). Similarly, \textbf{polysemanticity} in large models complicates interpretability-based auditing. In LLM-based agents, a single neuron or attention head may encode multiple unrelated concepts depending on context (\hyperlink{BereskaGavves2024}{Bereska \& Gavves, 2024}). An auditor might identify a coordination circuit, but suppressing it could inadvertently degrade the agent's ability to perform legitimate tasks.

Finally, monitoring and auditing also suffer from the \textbf{attribution problem}. Audits may identify collusive outcomes without cleanly attributing them to a specific agent, component, or training choice. This complicates governance -- who is sanctioned, retrained, or excluded (\hyperlink{KastnerEtAl2026}{Kästner et al., 2026}; \hyperlink{ZhangEtAl2025}{Zhang et al., 2025}; \hyperlink{Coeckelbergh2020}{Coeckelbergh, 2020})?

\subsection{Market Design \& Structural Measures}

While sanctions and monitoring react to collusion after it occurs, market design and structural measures aim to prevent it ex-ante by reshaping the interaction environment itself. When translated to multi-agent AI, the core idea is to design environments, protocols, and interaction topologies such that sustained undesired coordination is difficult to learn, maintain, or exploit.

\subsubsection{Implementation Approaches}

\begin{enumerate}
    \item \textbf{Interaction protocol design}\\
    The structure of agent interactions fundamentally determines collusion feasibility. Protocols that allow rich and transparent interaction (e.g., continuous price signaling, shared memory, persistent chat channels) are highly conducive to collusion (\hyperlink{MartyWarin2025}{Marty \& Warin, 2025}). Platforms can instead impose structurally collusion-hostile interaction protocols (\hyperlink{MarroEtAl2024}{Marro et al., 2024}). This includes one-shot interactions, where agents submit actions without observing contemporaneous actions of others (similar to sealed-bid auctions), thereby reducing the ability to signal intentions or punish deviations (\hyperlink{AgranovYariv2018}{Agranov \& Yariv, 2018}). Similarly, restricting communication schemas to task-relevant information and limiting or filtering free-form messaging reduces agents’ capacity to establish and sustain covert coordination strategies while preserving functionality necessary for legitimate task execution (\hyperlink{MotwaniEtAl2023PerfectCollusion}{Motwani et al., 2023}; \hyperlink{BonjourEtAl2022}{Bonjour et al., 2022}; \hyperlink{CalvanoEtAl2020}{Calvano et al.\ (2020)}).

    \item \textbf{Information Architecture}\\
    AI system designers can deliberately structure information flows to undermine collusion (\hyperlink{Zhang2025}{Zhang, 2025}; \hyperlink{BonjourEtAl2022}{Bonjour et al., 2022}; \hyperlink{MartyWarin2025}{Marty \& Warin, 2025}):
    \begin{itemize}
        \item \textbf{Differential information access:} Rather than providing all agents with identical environmental observations, systems can partition information such that each agent receives a different, partially overlapping view of the state space. This asymmetry makes it difficult for agents to verify whether coordination is being maintained by others, as discrepancies might reflect either defection or simply different information sets (\hyperlink{Zhang2025}{Zhang, 2025}; \hyperlink{MotwaniEtAl2023PerfectCollusion}{Motwani et al., 2023}).
        \item \textbf{Delayed or aggregated feedback:} Instead of providing immediate, granular feedback on competitor actions, systems can report only aggregated statistics with temporal delays. In LLM-based systems, this may involve preventing agents from accessing full conversation histories, shared memory across agents, or system-wide analytics. This increases uncertainty and weakens agents’ ability to monitor compliance with collusive strategies (\hyperlink{Zhang2025}{Zhang, 2025}).
        \item \textbf{Anonymization:} Systems can introduce anonymization or identity rotation schemes where agents interact with randomized or pseudonymous identifiers that change over time, preventing the formation of stable partnerships needed for collusion (\hyperlink{ShahriarEtAl2023}{Shahriar et al., 2023}; \hyperlink{MotwaniEtAl2023PerfectCollusion}{Motwani et al., 2023}).
    \end{itemize}

    \item \textbf{Agent population design}\\
    Collusion is easier when agents are homogeneous, repeatedly interact, and have aligned objectives (\hyperlink{DariusEtAl2025}{Darius et al., 2025}). A good approach to address this is entry facilitation as in human domains. Rather than maintaining a fixed set of agents, platforms can continuously introduce new agents, retire underperforming ones, or randomly shuffle participants across interaction groups. This constant churn prevents the stable conditions required for collusive equilibria. Even if some agents learn to coordinate, the frequent introduction of \textit{outsiders} who haven't learned the collusive strategy can destabilize the cartel. Another key approach is to introduce heterogeneity. In AI, a major risk factor is algorithmic monoculture -- if all agents are fine-tuned instances of the same foundation model (e.g., GPT-5.1), they are more likely to converge on a shared Schelling point, facilitating tacit collusion (\hyperlink{DariusEtAl2025}{Darius et al., 2025}; \hyperlink{BommasaniEtAl2022}{Bommasani et al., 2022}). Therefore, deploying agents with different architectures, training procedures, or foundation model providers increases diversity in learned strategies (\hyperlink{KeppoEtAl2025}{Keppo et al. 2025}; \hyperlink{DariusEtAl2025}{Darius et al., 2025}). For example, \hyperlink{KeppoEtAl2025}{Keppo et al. (2025}) show that a higher number of agents and algorithmic diversity reduces collusion among LLM agents.
\end{enumerate}

\subsubsection{Open Challenges}

Despite their promise, market design interventions for AI systems face several fundamental challenges. The primary challenge is the trade-off between anti-collusion design choices and system efficiency. Restricting communication may prevent coordination on collusion but also hinders legitimate cooperation on shared tasks. Moreover, AI agents may learn to exploit unintended channels created by the design itself. For instance, even when direct communication is restricted, agents may coordinate through timing, action ordering, or subtle correlations in outputs. Structural measures therefore risk being circumvented unless they are periodically revised or randomized (\hyperlink{GreenblattEtAl2023}{Greenblatt et al., 2023}; \hyperlink{BengioEtAl2025}{Bengio et al., 2025}; \hyperlink{HammondEtAl2025}{Hammond et al.\ (2025)}; \hyperlink{Ghaemi2025}{Ghaemi, 2025}).

Additionally, there is no universal anti-collusion architecture. The effectiveness of specific structural measures depends heavily on the domain, agent capabilities, training procedures, and task requirements. A design that prevents price collusion in a simulated marketplace might be ineffective or counterproductive in a multi-agent recommendation system or cooperative robotics setting. This context-dependence makes it difficult to establish general design principles or reusable architectural patterns. And the challenge intensifies for emerging AI applications, as we cannot confidently predict which structural features will prevent collusion because we don't yet understand what competitive behavior should look like or what coordination strategies might emerge. Early structural choices may inadvertently enable collusion in ways that only become apparent after large-scale deployment and are costly to correct (\hyperlink{Ghaemi2025}{Ghaemi, 2025}; (\hyperlink{DariusEtAl2025}{Darius et al., 2025}).

Furthermore, structural measures may impact other anti-collusion mechanisms proposed in this paper such as sanctions, monitoring, and leniency programs in counterproductive ways. For example, communication restrictions might make leniency programs less effective if agents cannot easily report collusion they are part of, or monitoring might become prohibitively expensive if architectural choices scatter evidence across opaque distributed systems.

Finally, market design and structural measures may advantage certain agents over others, such as agents better at operating under uncertainty, leading to unintended bias. Over time, this may concentrate influence among agents optimized for navigating restrictive environments, potentially recreating collusive dynamics in new forms. Therefore, optimal anti-collusion strategies likely require carefully balancing multiple complementary mechanisms rather than relying on structural measures alone.

\subsection{Governance}

Governance mechanisms for multi-agent AI operate at two distinct but interconnected layers: (1) human governance: policies, procedures, and organizational structures that shape how AI systems are developed, deployed, and overseen; and (2) system governance: automated architectural features and protocols embedded within the multi-agent AI itself. Effective anti-collusion governance requires combining both layers.

\subsubsection{Implementation Approaches}

\begin{enumerate}
    \item \textbf{Transparency and Documentation}\\
    Governance requires visibility into the "corporate DNA" of the agent. \hyperlink{MitchellEtAl2018}{Mitchell et al. (2018}) proposed Model Cards to standardize the documentation of model capabilities and limitations. Platforms can mandate comprehensive documentation of agent design, including model architecture specifications, training data sources, objective function definitions, and capability profiles, building on documentation practices such as datasheets for datasets (\hyperlink{GebruEtAl2018}{Gebru et al., 2018}). For open-weight models deployed by multiple parties, governance frameworks can require disclosure of fine-tuning procedures, adapter weights, and system prompts.

    \item \textbf{Separation of Oversight and Operation}\\
    A fundamental governance principle is preventing those being regulated from controlling the regulators. In AI systems, this maps to architectural separation between agent developers, deployment operators, monitoring and audit components, and policy-makers. For example, the team that fine-tunes agents for performance should not control collusion detection thresholds or sanction triggers. Aside from internal oversight, there should also be independent and external or oversight layers that periodically review system behavior, training updates, and interaction logs. This can include third-party audits of multi-agent deployments, red-teaming exercises focused explicitly on collusion scenarios, and mandatory external evaluations before scaling deployment (\hyperlink{Kolt2025}{Kolt, 2025}; \hyperlink{RajiEtAl2020}{Raji et al., 2020}; \hyperlink{PurpuraEtAl2025}{Purpura et al., 2025}).

    \item \textbf{Rotation policies}\\
    Human regulators often rotate staff to prevent cognitive capture where overseers adopt the assumptions or blind spots of regulated entities. In AI systems, this maps to rotating agents' model versions, interaction partners, or oversight configurations. Agents can be periodically re-seeded, retrained on fresh data, or reallocated across task pools, while oversight agents themselves can be rotated to prevent overfitting to known collusion patterns. These interventions reduce the stability required for collusive equilibria and limit agents’ ability to exploit fixed oversight structures over long horizons (\hyperlink{DingEtAl2024}{Ding et al., 2024}; \hyperlink{KeppoEtAl2025}{Keppo et al., 2025}).

    \item \textbf{Staged Deployment}\\
    Rather than deploying multi-agent AI systems directly into high-stakes environments, governance frameworks can mandate staged rollouts with escalating risk exposure. Initial deployments occur in simulated environments, allowing observation of emergent coordination before real-world consequences. Systems then progress to limited real-world pilots with restricted scope and reduced stakes. Only after demonstrating robust resistance to collusion across multiple test environments do systems advance to full deployment (\hyperlink{Kolt2025}{Kolt, 2025}; \hyperlink{RajiEtAl2020}{Raji et al., 2020}).

    \item \textbf{Shutdown}\\
    Automated governance has limits. Effective frameworks often require a "kill switch" operated by humans. If the monitoring layer (Section 3.3) raises a high-priority alert regarding systemic collusion, governance protocols must empower human operators to freeze the environment or suspend specific agents immediately. This prevents runaway feedback loops where algorithmic collusion spirals into a market crash or hyper-inflationary event before automated correctives can kick in (\hyperlink{Thornley2025}{Thornley, 2025}; \hyperlink{SchlatterEtAl2025}{Schlatter et al., 2025}).
\end{enumerate}

\subsubsection{Open Challenges}

Here, the primary challenge is automated governance at scale. Human governance involves relatively small numbers of decisions (thousands of transactions, dozens of regulated entities). Multi-agent AI systems may involve millions of interactions per day across thousands of autonomous agents. Human oversight cannot operate at this scale, yet fully automated governance creates the risk that oversight systems themselves develop blind spots or exploitable patterns. The challenge is designing governance that combines automated monitoring (for scale) with meaningful human review (for judgment and accountability) without creating bottlenecks (\hyperlink{Kolt2025}{Kolt, 2025}; \hyperlink{RajiEtAl2020}{Raji et al., 2020}).

Also, opacity and interpretability severely limit governance effectiveness. In deep neural networks, especially large language models, it is difficult to understand an agent's decision processes. Even with full access to model weights and activations, it may be impossible to determine why an agent took a specific action or whether coordination was intentional versus emergent (\hyperlink{SinghEtAl2024}{Singh et al., 2024}; \hyperlink{BereskaGavves2024}{Bereska \& Gavves, 2024}). This opacity undermines governance mechanisms that rely on intent assessment.

Meanwhile, the speed of adaptation in AI systems outpaces traditional governance cycles. Human regulatory frameworks evolve over years or decades, while AI agents can adapt strategies in hours or minutes (\hyperlink{FenwickEtAl2016}{Fenwick et al., 2016}). By the time governance protocols identify and address one collusion mechanism, agents may have evolved entirely new coordination strategies. This creates a perpetual lag problem where governance is always reactive, never proactive. Effective AI governance may require real-time adaptive mechanisms that adjust rules automatically based on observed behavior, but this raises concerns about legitimacy and accountability when rules change without human review (\hyperlink{Kolt2025}{Kolt, 2025}).

In addition, if governance systems themselves use AI components (e.g., automated monitoring), there is a risk of meta-level collusion where operational agents manipulate or corrupt oversight systems. Preventing this requires governance of the governance layer, creating recursive complexity.

The attribution and liability problem is another issue that is particularly severe for governance. In human systems, legal entities can be held accountable for collusion. In multi-agent AI, it is often unclear who should be held responsible -- the model developers who created the foundation model, those who fine-tuned and deployed agents, the platform operators who orchestrated interactions, or the agents themselves (which lack legal personhood) (\hyperlink{LiangEtAl2025}{Liang et al., 2025}; \hyperlink{ZhangEtAl2025}{Zhang et al., 2025}).

Finally, there's a fundamental trade-off between governance strength and innovation (\hyperlink{MartyWarin2025}{Marty \& Warin, 2025}). Stringent governance requirements impose substantial costs and delays that may inhibit beneficial AI development. Conversely, permissive governance accelerates innovation but increases risk. The optimal balance is unclear and likely varies by domain and risk level. Moreover, this trade-off creates competitive pressure -- jurisdictions or platforms with lighter governance may attract more development activity, potentially creating a race to the bottom in AI safety standards (\hyperlink{KashefiEtAl2024}{Kashefi et al., 2024}).

\section{Conclusion}

\subsection{Summary of Contributions}

This paper addresses a critical gap in multi-agent AI safety by systematically mapping human anti-collusion mechanisms to interventions for AI systems. First, we developed a taxonomy of five core anti-collusion mechanisms used in human domains: sanctions (penalties that reduce collusion payoffs), leniency and whistleblowing (mechanisms that destabilize cartels from within), monitoring and auditing (continuous observation and forensic examination), market design and structural measures (ex-ante interventions that make collusion difficult to sustain), and governance (institutional frameworks that foster integrity and limit discretion). For each mechanism, we identified representative tools and practices from high-stakes domains including financial markets, public procurement, telecommunications, and energy markets.

Second, we mapped each human mechanism to concrete implementation approaches for multi-agent AI. For sanctions, we outlined reward penalties, capability restrictions, and participation exclusions. For leniency and whistleblowing, we proposed self-reporting incentives, dedicated whistleblower agents, and steganographic disclosure rewards. For monitoring and auditing, we proposed telemetry-first system design, overseer agents, and triggered audits with replayability. For market design, we identified interaction protocol constraints, information architecture interventions, and agent population heterogeneity. And for governance, we identified transparency requirements, separation of oversight from operation, rotation policies, staged deployment, and circuit breakers as key implementation approaches.

Third, we identified fundamental open challenges that distinguish AI collusion from human collusion. These include the attribution problem (difficulty tracing collusive outcomes to specific network weights or training episodes), identity fluidity (agents can be forked or modified at near-zero cost, evading sanctions), and the boundary problem (distinguishing harmful collusion from beneficial cooperation and coordination from mere correlation). Other challenges include scalability and computational costs of comprehensive monitoring, steganography, adversarial adaptation creating arms races between detection and evasion, and the speed mismatch where AI agents adapt faster than governance protocols can respond.

\subsection{Limitations and Future Work}

This work has several important limitations that point toward future research directions. Our analysis is primarily conceptual and taxonomic rather than empirical. While we reference existing studies demonstrating AI collusion and preliminary interventions, we do not provide experimental validation of the full range of mechanisms we propose. Future work should systematically evaluate these interventions across diverse multi-agent AI environments, measuring both their effectiveness at reducing collusion and their impact on legitimate cooperation and system efficiency. Comparative studies assessing which mechanisms work best under what conditions would be particularly valuable.

Additionally, our framework treats each mechanism largely independently, but in practice these interventions interact in complex ways. Sanctions depend on monitoring for detection; leniency programs require verification infrastructure; market design choices affect what monitoring systems can observe; governance structures determine how all other mechanisms are implemented. Understanding these interactions, including potential synergies and conflicts, is critical for designing integrated anti-collusion systems. Research exploring optimal combinations and sequencing of mechanisms would advance both theory and practice.

Finally, we do not address the international and cross-jurisdictional dimensions of AI collusion governance. Multi-agent AI systems may operate across borders, raising questions about regulatory arbitrage, extraterritoriality, and coordination among national authorities. Research on global governance frameworks for multi-agent AI, drawing on international competition law and financial regulation, would complement our study.

\section*{References}
\begin{hangparas}{.5cm}{1}
\hypertarget{AskerNocke2021}{}Asker, J. W., \& Nocke, V. (2021). Collusion, mergers, and related antitrust issues. \emph{SSRN Electronic Journal}. \url{https://doi.org/10.2139/ssrn.3909620}

\hypertarget{CalvanoEtAl2020}{}Calvano, E., Calzolari, G., Denicolò, V., \& Pastorello, S. (2020). Artificial intelligence, algorithmic pricing, and collusion. \emph{American Economic Review}, 110(10), 3267--3297. \url{https://doi.org/10.1257/aer.20190623}

\hypertarget{CarboneEtAl2024}{}Carbone, C., Calderoni, F., \& Jofre, M. (2024). Bid-rigging in public procurement: cartel strategies and bidding patterns. \emph{Crime, Law, and Social Change}, 82(2), 249--281. \url{https://doi.org/10.1007/s10611-024-10142-0}

\hypertarget{ChassangOrtner2023}{}Chassang, S., \& Ortner, J. (2023). Regulating collusion. \emph{Annual Review of Economics}, 15(1), 177--204. \url{https://doi.org/10.1146/annurev-economics-051520-021936}

\hypertarget{ClarkEtAl2018}{}Clark, R., Coviello, D., Gauthier, J.-F., \& Shneyerov, A. (2018). Bid rigging and entry deterrence in public procurement: Evidence from an investigation into collusion and corruption in Quebec. \emph{Journal of Law, Economics, \& Organization}, 34(3), 301--363. \url{https://doi.org/10.1093/jleo/ewy011}

\hypertarget{IgamiSugaya2022}{}Igami, M., \& Sugaya, T. (2022). Measuring the incentive to collude: the vitamin cartels, 1990--99. \emph{The Review of Economic Studies}, 89, 1460--1494.

\hypertarget{HammondEtAl2025}{}Hammond, L., Chan, A., Clifton, J., Khan, A., McLean, E., Smith, C., Barfuss, W., Foerster, J., Gavenčiak, T., Han, T. A., Hughes, E., Kovařík, V., Kulveit, J., Leibo, J. Z., Oesterheld, C., De Witt, C. S., Shah, N., Wellman, M., Bova, P., \ldots{} Rahwan, I. (2025). Multi-Agent Risks from Advanced AI. \emph{arXiv preprint arXiv:2502.14143}. \url{https://arxiv.org/abs/2502.14143}

\hypertarget{MotwaniEtAl2024}{}Motwani, S. R., Baranchuk, M., Strohmeier, M., Bolina, V., Torr, P. H. S., Hammond, L., \& de Witt, C. S. (2024). Secret collusion among AI agents: Multi-agent deception via steganography. \emph{arXiv preprint arXiv:2402.07510}. \url{http://arxiv.org/abs/2402.07510}

\hypertarget{PawliczekEtAl2022}{}Pawliczek, A., Skinner, A. N., \& Zechman, S. L. C. (2022). Facilitating tacit collusion through voluntary disclosure: Evidence from common ownership. \emph{Journal of Accounting Research}. \url{https://doi.org/10.1111/1475-679x.12452}

\hypertarget{Symeonidis2018}{}Symeonidis, G. (2018). Collusion, profitability and welfare: Theory and evidence. \emph{Journal of Economic Behavior \& Organization}, 145, 530--545. \url{https://doi.org/10.1016/j.jebo.2017.11.010}

\hypertarget{ChicaEtAl2024}{}Chica, C., Guo, Y., \& Lerman, G. (2024). Artificial intelligence and algorithmic price collusion in two-sided markets. \emph{arXiv preprint arXiv:2407.04088}. \url{http://arxiv.org/abs/2407.04088}

\hypertarget{EpiventLambin2024}{}Epivent, A., \& Lambin, X. (2024). On algorithmic collusion and reward–punishment schemes. \emph{Economics Letters}, 237(111661), 111661. \url{https://doi.org/10.1016/j.econlet.2024.111661}

\hypertarget{Tailor2025}{}Tailor, O. (2025). Audit the Whisper: Detecting steganographic collusion in multi-agent LLMs. \emph{arXiv preprint arXiv:2510.04303}. \url{https://doi.org/10.48550/arXiv.2510.04303}

\hypertarget{PerreauDePinninck2010}{}Perreau de Pinninck, A., Sierra, C., \& Schorlemmer, M. (2010). A multiagent network for peer norm enforcement. \emph{Autonomous Agents and Multi-Agent Systems}, 21(3), 397--424. \url{https://doi.org/10.1007/s10458-009-9107-8}

\hypertarget{ECTrucks2016}{}European Commission. (2016). Antitrust: Commission fines truck producers 2.93 billion euros for participating in a cartel (Press Release IP/16/2582). \emph{European Commission}. \url{https://ec.europa.eu/commission/presscorner/detail/en/ip_16_2582}

\hypertarget{ECDeliveryHero2025}{}European Commission. (2025). Antitrust: Commission fines Delivery Hero and Glovo for participating in a cartel (Press Release IP/25/1356). \emph{European Commission}. \url{https://ec.europa.eu/commission/presscorner/detail/en/ip_25_1356}

\hypertarget{ACCCBingo2024}{}Australian Competition and Consumer Commission (ACCC) \& Commonwealth Director of Public Prosecutions (CDPP). (2024). Criminal sentences imposed on Bingo, Aussie Skips and their former CEOs Daniel Tartak and Emmanuel Roussakis for skip bin and waste processing cartel (Press release, February 2024). \href{https://www.accc.gov.au/media-release/criminal-sentences-imposed-on-bingo-aussie-skips-and-their-former-ceos-daniel-tartak-and-emmanuel-roussakis-for-skip-bin-and-waste-processing-cartel}{ACCC Press release}.

\hypertarget{WorldBankColas2022}{}World Bank. (2022). World Bank Group debars Colas Madagascar S.A. (Press release, 4 January 2022). \href{https://www.worldbank.org/en/news/press-release/2022/01/04/world-bank-group-debars-colas-madagascar-s-a}{World Bank press release - 2022}.

\hypertarget{WorldBankLSD2025}{}World Bank. (2025). World Bank Group debars L.S.D. Construction \& Supplies (Press release, 28 May 2025). \href{https://www.worldbank.org/en/news/press-release/2025/05/28/world-bank-group-debars-l-s-d-construction-supplies}{World Bank press release - 2025}.

\hypertarget{Miller2009}{}Miller, N. H. (2009). Strategic leniency and cartel enforcement. \emph{American Economic Review}, 99(3), 750--768. \url{https://doi.org/10.1257/aer.99.3.750}

\hypertarget{Directive2019Whistleblower}{}European Union. (2019). Directive (EU) 2019/1937 of the European Parliament and of the Council of 23 October 2019 on the protection of persons who report breaches of Union law. \emph{Official Journal of the European Union}, L 305, 17--56. \url{https://eur-lex.europa.eu/eli/dir/2019/1937/oj}

\hypertarget{ECWhistleblower2017}{}European Commission. (2017). Antitrust: Commission introduces new anonymous whistleblower tool (Press release IP/17/591, 16 March 2017). \emph{European Commission}. \url{https://europa.eu/rapid/press-release_IP-17-591_en.htm}

\hypertarget{NyrerodSpagnolo2021}{}Nyrer\"od, T., \& Spagnolo, G. (2021). A fresh look at whistleblower rewards. \emph{Journal of Governance and Regulation}, 10(4, special issue), 147--161. \url{https://doi.org/10.22495/jgrv10i4siart5}

\hypertarget{BrownEtAl2023ElectricityScreens}{}Brown, D. P., Eckert, A., \& Silveira, D. (2023). Screening for collusion in wholesale electricity markets: A literature review. \emph{Utilities Policy}, 85(101671), 101671. \url{https://doi.org/10.1016/j.jup.2023.101671}

\hypertarget{ProzHuber2025MLWholesaleElectricity}{}Proz, J., \& Huber, M. (2025). Machine learning for detecting collusion and capacity withholding in wholesale electricity markets. \emph{arXiv preprint arXiv:2508.09885}. \url{https://doi.org/10.48550/arXiv.2508.09885}

\hypertarget{GarciaRodriguezEtAl2022MLProcurement}{}García Rodríguez, M. J., Rodríguez-Montequín, V., Ballesteros-Pérez, P., Love, P. E. D., \& Signor, R. (2022). Collusion detection in public procurement auctions with machine learning algorithms. \emph{Automation in Construction}, 133(104047), 104047. \url{https://doi.org/10.1016/j.autcon.2021.104047}

\hypertarget{HuberImhofIshii2022TransnationalScreens}{}Huber, M., Imhof, D., \& Ishii, R. (2022). Transnational machine learning with screens for flagging bid-rigging cartels. \emph{Journal of the Royal Statistical Society: Series A (Statistics in Society)}, 185(3), 1074--1114. \url{https://doi.org/10.1111/rssa.12811}

\hypertarget{WallimannImhofHuber2023IncompleteCartels}{}Wallimann, H., Imhof, D., \& Huber, M. (2023). A machine learning approach for flagging incomplete bid-rigging cartels. \emph{Computational Economics}, 62(4), 1669--1720. \url{https://doi.org/10.1007/s10614-022-10315-w}

\hypertarget{HarringtonImhof2022CartelScreeningML}{}Harrington, J. E., Jr., \& Imhof, D. (2022). Cartel screening and machine learning. \emph{Stanford Computational Antitrust}, 2, 133--170.

\hypertarget{DusoEtAl2025PublicCommunicationCollusion}{}Duso, T., Harrington, J. E., Jr., Kreuzberg, C., \& Sapi, G. (2025). Public communication and collusion: New screening tools for competition authorities. \emph{SSRN Working Paper}. \url{https://doi.org/10.2139/ssrn.5373701}

\hypertarget{ECShipping2016}{}European Commission. (2016). Antitrust: Commission accepts commitments from container liner shipping companies on price transparency (Press Release IP/16/2446). \emph{European Commission}. \url{https://ec.europa.eu/commission/presscorner/detail/en/IP_16_2446}

\hypertarget{Klemperer2002}{}Klemperer, P. (2002). What really matters in auction design. \emph{The Journal of Economic Perspectives}, 16(1), 169–189. \url{https://doi.org/10.1257/0895330027166}

\hypertarget{BourreauEtAl2021}{}Bourreau, M., Sun, Y., \& Verboven, F. (2021). Market entry, fighting brands, and tacit collusion: Evidence from the French mobile telecommunications market. \emph{American Economic Review}, 111(11), 3459--3499. \url{https://doi.org/10.1257/aer.20190540}

\hypertarget{Rilinger2023}{}Rilinger, G. (2023). Who captures whom? Regulatory misperceptions and the timing of cognitive capture. \emph{Regulation \& Governance}, 17(1), 43--60. \url{https://doi.org/10.1111/rego.12438}

\hypertarget{YeEtAl2025}{}Ye, K., Yaoliang, Z., Zhenhua, H., \& Bingzhen, L. (2025). Collusive bidding in the construction industry: a systematic review of organizational forms, mechanisms and anti-collusion strategies. \emph{Building Research \& Information}, 1--15. \url{https://doi.org/10.1080/09613218.2025.2574682}

\hypertarget{Reyes2023}{}Reyes, E. A. (2023). Enhancing the impact of Integrity Pacts in public procurement (Working Paper No. 46). \emph{Basel Institute on Governance}.

\hypertarget{CarpenterMoss2014}{}Carpenter, D., \& Moss, D. A. (2014). \emph{Preventing Regulatory Capture: Special Interest Influence and How to Limit It}. Cambridge University Press. \url{https://doi.org/10.1017/CBO9781139565875}

\hypertarget{Coeckelbergh2020}{}Coeckelbergh, M. (2020). Artificial intelligence, responsibility attribution, and a relational justification of explainability. \emph{Science and Engineering Ethics}, 26(4), 2051--2068. \url{https://doi.org/10.1007/s11948-019-00146-8}

\hypertarget{Ghaemi2025}{}Ghaemi, M. S. (2025). A survey of collusion risk in LLM-powered multi-agent systems. In \emph{Socially Responsible and Trustworthy Foundation Models at NeurIPS}.

\hypertarget{KastnerEtAl2026}{}Kästner, L., Cordes, J., \& Zech, H. (2026). Responsibility attribution for AI-mediated damages with mechanistic interpretability. In \emph{Lecture Notes in Computer Science} (pp. 187--202). Springer Nature Switzerland.

\hypertarget{TadimallaMaher2024}{}Tadimalla, S. Y., \& Maher, M. L. (2024). AI and identity. \emph{arXiv preprint arXiv:2403.07924}. \url{https://doi.org/10.48550/ARXIV.2403.07924}

\hypertarget{WuEtAl2024}{}Wu, J., Wu, Z., Xue, Y., Wen, J., \& Peng, W. (2024). Generative text steganography with large language model. \emph{Proceedings of the 32nd ACM International Conference on Multimedia}, 10345--10353.

\hypertarget{ZhangEtAl2025}{}Zhang, S., Du, H., Ma, J. W., \& Lakkaraju, H. (2025). Who gets credit or blame? Attributing accountability in modern AI systems. \emph{arXiv preprint arXiv:2506.00175}. \url{http://arxiv.org/abs/2506.00175}

\hypertarget{Banerjee2023}{}Banerjee, S. (2023). Combating algorithmic collusion: A mechanism design approach. \emph{arXiv preprint arXiv:2303.02576}. \url{http://arxiv.org/abs/2303.02576}

\hypertarget{ChenEtAl2023FireAct}{}Chen, B., Shu, C., Shareghi, E., Collier, N., Narasimhan, K., \& Yao, S. (2023). FireAct: Toward language agent fine-tuning. \emph{arXiv preprint arXiv:2310.05915}. \url{https://arxiv.org/abs/2310.05915}
\end{hangparas}

\hypertarget{Caviola2025}{}Caviola, L. (2025). The societal response to potentially sentient AI. \emph{arXiv preprint arXiv:2502.00388}. \url{http://arxiv.org/abs/2502.00388}

\hypertarget{ParkEtAl2024}{}Park, P. S., Goldstein, S., O’Gara, A., Chen, M., \& Hendrycks, D. (2024). AI deception: A survey of examples, risks, and potential solutions. \emph{Patterns}, 5(5), 100988. \url{https://doi.org/10.1016/j.patter.2024.100988}

\hypertarget{SinghEtAl2024}{}Singh, C., Inala, J. P., Galley, M., Caruana, R., \& Gao, J. (2024). Rethinking interpretability in the era of large language models. \emph{arXiv preprint arXiv:2402.01761}. \url{http://arxiv.org/abs/2402.01761}

\hypertarget{BengioEtAl2025}{}Bengio, Y., Cohen, M., Fornasiere, D., Ghosn, J., Greiner, P., MacDermott, M., Mindermann, S., Oberman, A., Richardson, J., Richardson, O., Rondeau, M.-A., St-Charles, P.-L., \& Williams-King, D. (2025). Superintelligent agents pose catastrophic risks: Can Scientist AI offer a safer path? \emph{arXiv preprint arXiv:2502.15657}. \url{http://arxiv.org/abs/2502.15657}

\hypertarget{deWitt2025}{}de Witt, C. S. (2025). Open challenges in multi-agent security: Towards secure systems of interacting AI agents. \emph{arXiv preprint arXiv:2505.02077}. \url{http://arxiv.org/abs/2505.02077}

\hypertarget{Musolff2022}{}Musolff, L. (2022). Algorithmic pricing facilitates tacit collusion: Evidence from e-commerce. In \emph{Proceedings of the 23rd ACM Conference on Economics and Computation} (pp. 32--33).

\hypertarget{SongEtAl2024}{}Song, C., Ma, L., Zheng, J., Liao, J., Kuang, H., \& Yang, L. (2024). Audit-LLM: Multi-agent collaboration for log-based insider threat detection. \emph{arXiv preprint arXiv:2408.08902}. \url{https://arxiv.org/abs/2408.08902}

\hypertarget{MokanderEtAl2024}{}Mökander, J., Schuett, J., Kirk, H. R., \& Floridi, L. (2024). Auditing large language models: a three-layered approach. \emph{AI and Ethics}, 4(4), 1085--1115. \url{https://doi.org/10.1007/s43681-023-00289-2}

\hypertarget{BereskaGavves2024}{}Bereska, L., \& Gavves, E. (2024). Mechanistic interpretability for AI safety--a review. \emph{arXiv preprint arXiv:2404.14082}. \url{https://arxiv.org/abs/2404.14082}

\hypertarget{PavlovaEtAl2024}{}Pavlova, M., Brinkman, E., Iyer, K., Albiero, V., Bitton, J., Nguyen, H., Li, J., Ferrer, C. C., Evtimov, I., \& Grattafiori, A. (2024). Automated red teaming with GOAT: The Generative Offensive Agent Tester. \emph{arXiv preprint arXiv:2408.01353}. \url{https://arxiv.org/abs/2408.01353}

\hypertarget{PerezEtAl2022}{}Perez, E., Huang, S., Song, F., Cai, T., Ring, R., Aslanides, J., Glaese, A., McAleese, N., \& Irving, G. (2022). Red teaming language models with language models. In \emph{Proceedings of the 2022 Conference on Empirical Methods in Natural Language Processing} (pp. 3419--3448). \url{https://doi.org/10.18653/v1/2022.emnlp-main.225}

\hypertarget{HuaEtAl2025}{}Hua, T. T., Baskerville, J., Lemoine, H., Hopman, M., Bhatt, A., \& Tracy, T. (2025). Combining cost-constrained runtime monitors for AI safety. \emph{arXiv preprint arXiv:2507.15886}. \url{https://arxiv.org/abs/2507.15886}

\hypertarget{WenEtAl2023}{}Wen, J., Zhang, Z., Lan, Y., Cui, Z., Cai, J., \& Zhang, W. (2023). A survey on federated learning: challenges and applications. \emph{International Journal of Machine Learning and Cybernetics}, 14(2), 513--535. \url{https://doi.org/10.1007/s13042-022-01647-y}

\hypertarget{KeppoEtAl2025}{}Keppo, J., Li, Y., Tsoukalas, G., \& Yuan, N. (2025). AI pricing, agent heterogeneity, and collusion. \emph{SSRN Electronic Journal}. \url{http://dx.doi.org/10.2139/ssrn.5386338}

\hypertarget{BonjourEtAl2022}{}Bonjour, T., Aggarwal, V., \& Bhargava, B. (2022). Information theoretic approach to detect collusion in multi-agent games. In \emph{Uncertainty in Artificial Intelligence} (pp. 223--232). PMLR.

\hypertarget{AgranovYariv2018}{}Agranov, M., \& Yariv, L. (2018). Collusion through communication in auctions. \emph{Games and Economic Behavior}, 107, 93--108. \url{https://doi.org/10.1016/j.geb.2017.10.021}

\hypertarget{BommasaniEtAl2022}{}Bommasani, R., Creel, K. A., Kumar, A., Jurafsky, D., \& Liang, P. (2022). Picking on the same person: Does algorithmic monoculture lead to outcome homogenization? \emph{arXiv preprint arXiv:2211.13972}. \url{http://arxiv.org/abs/2211.13972}

\hypertarget{DariusEtAl2025}{}Darius, P., Hoppe, T., \& Aleksandrov, A. (2025). Systemic risks of interacting AI. \emph{arXiv preprint arXiv:2512.17793}. \url{https://doi.org/10.48550/arXiv.2512.17793}

\hypertarget{MarroEtAl2024}{}Marro, S., La Malfa, E., Wright, J., Li, G., Shadbolt, N., Wooldridge, M., \& Torr, P. (2024). A scalable communication protocol for networks of large language models. \emph{arXiv preprint arXiv:2410.11905}. \url{http://arxiv.org/abs/2410.11905}

\hypertarget{MartyWarin2025}{}Marty, F., \& Warin, T. (2025). Deciphering algorithmic collusion: Insights from bandit algorithms and implications for antitrust enforcement. \emph{Journal of Economy and Technology}, 3, 34--43. \url{https://doi.org/10.1016/j.ject.2024.10.001}

\hypertarget{RenEtAl2025}{}Ren, S., Fu, W., Zou, X., Shen, C., Cai, Y., Chu, C., Wang, Z., \& Hu, S. (2025). Beyond the tragedy of the commons: Building a reputation system for generative multi-agent systems. \emph{arXiv preprint arXiv:2505.05029}. \url{http://arxiv.org/abs/2505.05029}

\hypertarget{Zhang2025}{}Zhang, N. (2025). Too noisy to collude? Algorithmic collusion under Laplacian noise. \emph{arXiv preprint arXiv:2509.02800}. \url{http://arxiv.org/abs/2509.02800}

\hypertarget{MotwaniEtAl2023PerfectCollusion}{}Motwani, S. R., Baranchuk, M., Hammond, L., \& de Witt, C. S. (2023). A perfect collusion benchmark: How can AI agents be prevented from colluding with information-theoretic undetectability? In \emph{Multi-Agent Security Workshop @ NeurIPS'23}.

\hypertarget{ShahriarEtAl2023}{}Shahriar, S., Allana, S., Hazratifard, S. M., \& Dara, R. (2023). A survey of privacy risks and mitigation strategies in the artificial intelligence life cycle. \emph{IEEE Access}, 11, 61829--61854. \url{https://doi.org/10.1109/ACCESS.2023.3286453}

\hypertarget{MathewEtAl2025}{}Mathew, Y., Matthews, O., McCarthy, R., Velja, J., de Witt, C. S., Cope, D., \& Schoots, N. (2025). Hidden in plain text: Emergence \& mitigation of steganographic collusion in LLMs. \emph{arXiv preprint arXiv:2410.03768}. \url{https://doi.org/10.48550/arXiv.2410.03768}

\hypertarget{GreenblattEtAl2023}{}Greenblatt, R., Shlegeris, B., Sachan, K., \& Roger, F. (2023). AI control: Improving safety despite intentional subversion. \emph{arXiv preprint arXiv:2312.06942}. \url{http://arxiv.org/abs/2312.06942}

\hypertarget{MitchellEtAl2018}{}Mitchell, M., Wu, S., Zaldivar, A., Barnes, P., Vasserman, L., Hutchinson, B., Spitzer, E., Raji, I. D., \& Gebru, T. (2018). Model cards for model reporting. \emph{arXiv preprint arXiv:1810.03993}. \url{http://arxiv.org/abs/1810.03993}

\hypertarget{GebruEtAl2018}{}Gebru, T., Morgenstern, J., Vecchione, B., Vaughan, J. W., Wallach, H., Daumé, H., III, \& Crawford, K. (2018). Datasheets for datasets. \emph{arXiv preprint arXiv:1803.09010}. \url{http://arxiv.org/abs/1803.09010}

\hypertarget{RajiEtAl2020}{}Raji, I. D., Smart, A., White, R. N., Mitchell, M., Gebru, T., Hutchinson, B., Smith-Loud, J., Theron, D., \& Barnes, P. (2020). Closing the AI accountability gap: Defining an end-to-end framework for internal algorithmic auditing. \emph{arXiv preprint arXiv:2001.00973}. \url{http://arxiv.org/abs/2001.00973}

\hypertarget{PurpuraEtAl2025}{}Purpura, A., Wadhwa, S., Zymet, J., Gupta, A., Luo, A., Rad, M. K., Shinde, S., \& Sorower, M. S. (2025). Building safe GenAI applications: An end-to-end overview of red teaming for Large Language Models. \emph{arXiv preprint arXiv:2503.01742}. \url{http://arxiv.org/abs/2503.01742}

\hypertarget{DingEtAl2024}{}Ding, W., Bhagoji, A. N., Zhao, B. Y., \& Zheng, H. (2024). Towards scalable and robust model versioning. In \emph{2024 IEEE Conference on Secure and Trustworthy Machine Learning (SaTML)}.

\hypertarget{Kolt2025}{}Kolt, N. (2025). Governing AI agents. \emph{arXiv preprint arXiv:2501.07913}. \url{http://arxiv.org/abs/2501.07913}

\hypertarget{Thornley2025}{}Thornley, E. (2025). The shutdown problem: an AI engineering puzzle for decision theorists. \emph{Philosophical Studies}, 182(7), 1653--1680. \url{https://doi.org/10.1007/s11098-024-02153-3}

\hypertarget{SchlatterEtAl2025}{}Schlatter, J., Weinstein-Raun, B., \& Ladish, J. (2025). Shutdown resistance in large language models. \emph{arXiv preprint arXiv:2509.14260}. \url{http://arxiv.org/abs/2509.14260}

\hypertarget{LiangEtAl2025}{}Liang, Z., Li, Y., \& Chen, T. (2025). When AI fails, who gets the blame? Citizens’ attribution patterns in AI-induced public service failures. \emph{Journal of Chinese Governance}, 10(4), 630--659. \url{https://doi.org/10.1080/23812346.2025.2528005}

\hypertarget{KashefiEtAl2024}{}Kashefi, P., Kashefi, Y., \& Ghafouri Mirsaraei, A. (2024). Shaping the future of AI: balancing innovation and ethics in global regulation. \emph{Uniform Law Review}, unae040. \url{https://doi.org/10.1093/ulr/unae040}

\hypertarget{FenwickEtAl2016}{}Fenwick, M., Kaal, W. A., \& Vermeulen, E. P. (2016). Regulation tomorrow: what happens when technology is faster than the law. \emph{American University Business Law Review}, 6, 561.
\end{document}